\documentstyle[aps,prd]{revtex}
\begin{document}

\newcommand{\be}{\begin{equation}}
\newcommand{\ee}{\end{equation}}
\newcommand{\ben}{\begin{eqnarray}}
\newcommand{\een}{\end{eqnarray}}
\newcommand{\nn}{// \nonumber}
\newcommand{\no}{\noindent}
\newcommand{\n}{\label}

\title{Inhomogeneous cosmologies with Q-matter and varying $\Lambda$}

\author{
Luis P. Chimento\footnote{Electronic address:
chimento@df.uba.ar}  and  Alejandro S. Jakubi\footnote{
Electronic address: jakubi@df.uba.ar}}
\address{Departamento de F\'{\i}sica, Universidad de 
Buenos Aires, 1428~Buenos Aires, Argentina}
\author{Diego Pav\'{o}n\footnote{Electronic address:
diego@ulises.uab.es}}
\address{
Departament de F\'{\i}sica, Universidad Aut\'onoma 
de Barcelona, 08193 Bellaterra, Spain}

\date{\today}

\maketitle


\begin{abstract}
Starting from the inhomogeneous shear--free Nariai metric we show,
by solving the Einstein--Klein--Gordon field equations,
how a self--interacting scalar field plus a material fluid, 
a variable cosmological term and a heat flux can
drive the universe to its currently observed state of homogeneous 
accelerated expansion. A quintessence scenario where power-law inflation 
takes place for a string-motivated potential in the late--time 
dominated field regime is proposed.
\end{abstract}

\section{Introduction}

The theory of general relativity implies that the large scale 
geometry and evolution of the Universe is dictated by
its content of matter and fields, including energy fluxes, 
shear stresses, particle production, and so on.
Recently, there have been claims in the literature 
that the Universe, besides its content in normal matter 
and radiation, must possess a not yet identified 
component (usually called {\em quintessence} matter, 
Q-matter for short) \cite{turner1}, \cite{caldwell}, 
\cite{zlatev}, \cite{bucher}, characterized 
by a negative pressure, and possibly a cosmological term 
which may be constant or not \cite{peebles}, 
\cite{turner2}. These claims were prompted 
at the realization that the clustered matter 
component can account at most for one third of the critical density.
Therefore, an additional ``soft" (i.e. non-clustered) component is 
needed if the critical density predicted by many inflationary 
models is to be achieved.

Very often the geometry of the proposed models is very simple,
just Friedmann-Lema\^{\i}tre-Robertson-Walker (FLRW), partly 
because of mathematical simplicity, and partly because they 
deal with
not very early stages of cosmic evolution. However, up to now no
convincing argument has been avanced to the effect that the 
geometry of the very 
early Universe (say, some time before nucleosynthesis) could
not have been either anisotropic or inhomogeneous, or both.
Note that it is very natural to assume the geometry at that
primeval epoch more general than just FLRW.
Moreover, recently it has been demonstrated that given any
spherically symmetric geometry and any set of observations,
evolution functions for the sources can be found that will
make the model compatible in general with 
observation \cite{mustapha}.  
Further motivations to study the evolution of inhomogeneous 
cosmological models can be found in references
\cite{maccallum} and \cite{krasinski}. 
In constrast to FLRW models, inhomogeneous
spaces are in general compatible with heat fluxes, and these might 
imply important consequences such as inflation \cite{roy} or the
avoidance of the initial singularity \cite{dadhich}. Here 
we focus on an isotropic but inhomogeneous spherically symmetric
universe which besides a material fluid contains a 
self-interacting scalar field (which can be interpreted 
as Q-matter), and a cosmological term, $\Lambda$ which, 
in general, may vary with time. As it turns out the 
homogeneization of the Universe at large times  depends
very much on the value the effective adiabatic index
$\gamma$, defined below (Eqn. (\ref{gamma})), takes. When 
more generally this quantity varies with
time, homogeneization can also be achieved under 
rather ample conditions. 

Obviously, density inhomogenities triggered by gravitational
instability must be present at any stage of evolution, 
even when the universe has reached a near homogeneous state 
describable in the mean by the FLRW metric -i.e. an average 
homogeneous universe housing growing as well as decaying 
density perturbation modes, the former leading eventually to 
the formation of the cosmic structures we observe today. 
About the evolution of scalar (i.e. density) inhomogeneities there 
is a rather large and still growing body of literature (see {\em e.g.}
\cite{efs} and references therein) but we shall not deal with them
as they lie outside the main focus of this paper. We only mention
that the negative pressure associated to  Q-matter and 
$\Lambda$ will tend to slow down the growing modes
(see {\em e.g.} \cite{turner1}, \cite{efs}, \cite{adz}),
and shift the epoch of matter-radiation equality toward 
more recent times \cite{cbfp}. A detailed study of all 
this will be the subject of a future work.
 
Section II  presents the basic equations. Section III
solves the Einstein-Klein-Gordon (EKG) field equations 
assuming $\gamma$ (but not $\Lambda$) a constant. 
Section IV studies the asymptotic evolution toward a 
Q-matter dominated era. Section  V solves 
the EKG equations when $\gamma$ 
is a function of time. Finally, section VI
summarizes the findings of this paper. 
Units have been chosen so that $c = G = 1$.

\section{Einstein-Klein-Gordon field equations}

Let us consider a shear--free spherically--symmetric 
spacetime with metric \cite{nariai}

\[
{ \ ds}^{2} = - {A}(t, r)^{2} {\ d}\,t^{2} + {B}(t, r)^{2}
\left[ {\ d}\, r^{2} + r^{2} \; \mbox{d}\Omega^{2}\right]  \, ,
\]
where as usual
$\mbox{d}\Omega^2 \equiv \mbox{d}\theta^{2} + \sin^{2}\, \theta \,
\mbox{d} \phi^{2}$. For later purposes we introduce the 
functions ${F}(t,r) \equiv 1/{B}(t, r)$ and 
$ v(t,r) \equiv A F$. Thus, the line element assumes the
more convenient form  
\begin{equation}\label{ds3}
{ \ ds}^{2}=\frac{1}{{ F}(t, \,r)^{2}}\left[
\,
- v(t,r)^{2}\,{ \ d}\,t^{{ 2\ }}+
 \ d\,r^2  +
  r^2\, \mbox{d}\Omega^2
    \right].
\end{equation}

As sources of the gravitational field we take: a fluid of material
energy density $\rho_f = \rho_f (r,t)$, hydrostatic pressure 
$p_f = p_f(r,t)$, with a radial heat flow 
($q_{r} = q_{r}(r,t) \, $ and $\, q_{t} = q_{\theta} = 
q_{\phi} = 0$), plus a cosmological term,
related to the energy density of vacuum by 
$\Lambda = 8\pi \rho_{vac}$, 
that depends only on time $\Lambda = \Lambda(t)$, and a
self-interacting scalar field $\phi$ driven by the potential
$V(\phi)$.
Taking the scalar field to depend only on $t$, its
energy-momentum tensor
 may be
written in the perfect fluid form

\begin{equation}\label{Tfluid}
T_{ik}= (p_{\phi}+\rho_{\phi}) u_{i}u_{k}+p_{\phi}g_{ik}\,,
\end{equation}

\noindent
where
$u^i=\phi^{,i}/\sqrt{-\phi_{,l}\phi^{,l}}$
together with
\begin{eqnarray}\label{prhophi}
\rho_{\phi} & = &  -\frac{1}{2} \phi_{,l}\phi^{,l} + V(\phi)\, ,
\nonumber \\
   p_{\phi} & = &  -\frac{1}{2} \phi_{,l}\phi^{,l} - V(\phi).
\end{eqnarray}

\noindent
The fluid interpretation of the scalar field,
while not compulsory, has proven very useful in the 
study of the inflationary phase and reheating of the universe 
-see for instance \cite{zd1} and \cite{zd2}. In particular it leads to
consider its equation of state
$p_{\phi}=\left(\gamma_{\phi}-1\right)\rho_{\phi}$.
Hence the scalar
field can be interpreted as Q-matter since, depending on $V(\phi)$,
$\gamma_{\phi}$ may be lower than one -see e.g.
\cite{zlatev}. The stress energy-tensor of the normal matter, 
with a heat flow, plus Q-matter (scalar field) and the 
cosmological term is 
\begin{equation}
T^{i}_{k} = (\rho_f+\rho_{\phi} + p_f+p_{\phi}) u^{i} u_{k} +
(\Lambda - p_f-p_{\phi}) \delta^{i}_{k} + q^{i} u_{k}
+q_{k} u^{i} \, ,
\label{2}
\end{equation}
where the comoving four-velocity is normalized so that $u^{i} u_{i} = - 1$,
with $u^{i} = \delta^{i}_{t} (g_{tt})^{- 1/2}$, and hence
$u^{t} = F/v$.
Obviously the heat flow is orthogonal to
$u_{i}\, $ (i.e. $q^{i} u_{i} = 0$).
As equation of state for the fluid we choose
$p_f=\left(\gamma_f-1\right)\rho_f$
where $\gamma_{f}$ is a function of $t$ and $r$.
Taking into account the additivity of the
stress-energy tensor it makes sense to consider an effective perfect fluid
description with equation of state $p=\left(\gamma-1\right)\rho$ where
$p=p_f+p_{\phi}$, $\rho=\rho_f+\rho_{\phi}$ and

\be
\n{gamma}
\gamma=\frac{\gamma_f\rho_f+\gamma_{\phi}\rho_{\phi}}
{\rho_f+\rho_{\phi}} \, ,
\ee
is the overall (i.e. effective) adiabatic index.

The requirement that the cosmological term $\Lambda$ is 
just a function of $t$ leads to the restriction
that $\gamma$ also depends only on $t$ to render the system 
of Einstein equations integrable. The nice result we are seeking 
is a solution that has an asymptotic FLRW stage,
with $\Lambda $ evolving towards a constant, and 
the heat flow vanishing in that limit.

For the metric (\ref{ds3})
the EKG equations are

\begin{equation}\label{00}
12\,x\,{ F^{\prime }}^{2} - 12\,{ F}\,
{ F^{\prime }} - 8\,x{ F}\,{ F^{\prime \prime }} -
3\,{\displaystyle \frac {{ \dot F}^{2}}{{ v}
^{2}}}  + \rho  + \Lambda=0,
\end{equation}

$$
 12\,x\,{ F^{\prime }}^{2} - 8\,{ F}
 \,{ F^{\prime }} - 3\,{\displaystyle \frac {{ \dot F}^{2}
}{{ v}^{2}}}  - 2\,{\displaystyle \frac {{ F}
\,{ \dot F}\,{ \dot v}}{{ v}^{3}}}  + 2\,
{\displaystyle \frac {{ F}\,{ \ddot F}}{{ v}
^{2}}}  + 4\,{\displaystyle \frac {{ F}^{2}\,{ v
^{\prime }}}{{ v}}}
$$
\begin{equation} \label{11}
 - 8\,{\displaystyle \frac {x{ F}\,{
F^{\prime }}\,{ v^{\prime }}}{{ v}}}  - p + \Lambda=0,
\end{equation}

$$
 12\,x\,{ F^{\prime }}^{2} - 8\,{ F}
 \,{ F^{\prime }} - 8\,x{ F}\,{ F
^{\prime \prime }} - 3\,{\displaystyle \frac {{ \dot F}^{2}}{
{ v}^{2}}}  - 2\,{\displaystyle \frac {{ F}
\,{ \dot F}\,{ \dot v}}{{ v}^{3}}}
$$
\begin{equation} \label{22}
+ 2\,{\displaystyle \frac {{ F}\,{ \ddot F
}}{{ v}^{2}}}  - 8\,{\displaystyle \frac {x{ F}
\,{ F^{\prime }}\,{ v^{\prime }}}{{ v}}}
 + 4\,{\displaystyle \frac {{ F}^{2}\,{ v^{\prime }
}}{{ v}}}  + 4\,{\displaystyle \frac {x{ F}
^{2}\,{ v^{\prime \prime }}}{{ v}}}  - p +
\Lambda=0,
\end{equation}

\begin{equation}\label{03}
 - 8\,{\displaystyle \frac {x{ F}^{2}\,
{ \dot F}\,{ v^{\prime }}}{{ v}^{2}}}  + 8\,
{\displaystyle \frac {x{ F}^{2}\,{ \dot F^{\prime }}
}{{ v}}}  + { q_{r}}=0 ,
\end{equation}

\begin{equation} \label{KG}
\ddot \phi-2\frac{\dot F}{F}\dot\phi+\frac{1}{F^2}\frac{dV}{d\phi}=0,
\end{equation}

\noindent
where $x=r^2$, an over-dot means
$\partial/\partial t$, and a prime 
$\partial/\partial x$. We thus have five equations and 
seven unknowns ($F \, , v \, ,\rho \, , p, \, \phi,\, q_{r} \,$  
and $\Lambda$). Hence two more equations are needed to
render the system determinate. These are the above 
equation of state relating $\rho$ and $p$ and the condition 
$\partial \Lambda/\partial x =0$.

{}From (\ref{11}) and (\ref{22}) we get

\begin{equation}\label{11-22}
\frac{v''}{v}=2\frac{F''}{F},
\end{equation}

\noindent
then using the ansatz $F=\left[a(t) + b(t)x\right]^{k}$ and
$v=\left[c(t) +d(t)x\right]^{n}$
in (\ref{11-22}) we find that the constraints
$n\left(n-1\right)=2k\left(k-1\right)$ and
$a(t)\, d(t)=b(t)\, c(t)$
must be satisfied.
Next we study the simplest case, namely $n=k=1$. This is the
most interesting instance because the functions $a(t), b(t), c(t)$ and $d(t)$
can be chosen freely. Another set of solutions can be obtained if 
the relationship $v_{xx}/v=2[\nu(t)]^2$ is assumed in (\ref{11-22}).
These solutions are presently under study and we shall
report on them in a future paper.

\subsection{Einstein equations with time-dependent $\Lambda$ and $\gamma$ }

When $n=k=1$ we obtain a set of solutions that contains those of Modak
($b=0$) \cite{modak}, Bergmann ($c=a, d=b$) \cite{bergmann}
and Maiti ($b=d=k \, a/4 \, ,$ with $k=0, \pm1$) \cite{maiti}.
Another possibility arises when $d=0$. This solution can be also obtained
from the metric (\ref{ds3}) by imposing this metric be conformal to
Minkowski's. Namely, the integrability of the transformation $-v dt+dr=d\eta$
and $v dt+dr=d\sigma$ leads to $v=v(t)$ and $n=k=1$. This allows us to
re-define the time by $v dt\to dt$. Then the metric (\ref{ds3}) becomes

\begin{equation}\label{ds}
{ \ ds}^{2}=\frac{1}{\left[a(t)+b(t)r^{2}\right]^2}\left(
\, - { \ d}\,t^{{ 2\ }} + \ d\,r^2  + r^2\, \mbox{d}\Omega^2 \right) ,
\end{equation}
and equations (\ref{00}), (\ref{11}) and (\ref{03})
turn into

\begin{equation} \label{00c}
\rho+\Lambda=12 ab+3\dot a^2+6\dot a\dot b x+3\dot b^2 x^2 ,
\end{equation}

\begin{equation} \label{11c}
p-\Lambda=\left(2b\ddot b-3\dot b^2\right)x^2+
2\left(2b^2-3\dot a\dot b+a\ddot b+\ddot a b\right)x-
8ab-3\dot a^2+2a \ddot{a} \, ,
\end{equation}
and
\begin{equation} \label{03c}
q_{r} =-4\sqrt{x}\dot b\left(a+bx\right)^{2} ,
\end{equation}
respectively.
We next impose  that $\Lambda$ depends solely on time,
so it must have the form

\begin{equation} \label{lambdact}
\Lambda(t) = 12 ab + 3 \dot{a}^2 + f(t) ,
\end{equation}
where $f(t)$ is a function to be determined. Then, (\ref{00c})
and (\ref{11c}) imply

\begin{equation} \label{rhoct}
\rho(x,t)=6\dot a\dot b x+3\dot b^2 x^2-f(t),
\end{equation}

\begin{equation} \label{pct}
p(x,t)=\left(2b\ddot b-3\dot b^2\right)x^2+
2\left(2b^2-3\dot a\dot b+a\ddot b+\ddot a b\right)x+
4ab+2a\ddot a+f(t).
\end{equation}
We further impose the equation of state $p=\left(\gamma-1\right)\rho$. Taking
into account that $a$ and $b$ depend only on time, and equating
the coefficients of same power of $x$ we obtain a set of 
equations to determine $a$, $b$ and $f$

\begin{equation} \label{x^2}
b\ddot b-\frac{3}{2}\gamma\dot b^2=0,
\end{equation}

\begin{equation} \label{x^1}
\ddot a-3\gamma\frac{\dot b}{b}\dot a+\frac{3\gamma}{2}\frac{\dot b^2}{b^2}a
=-2b\, ,
\end{equation}

\begin{equation} \label{x^0}
f=-\frac{2a}{\gamma}\left(2b+\ddot a\right).
\end{equation}

To show the conditions that lead asymptotically to a 
FLRW metric it is expedient to introduce
the time coordinate $d\tau= dt/a$. Thus (\ref{ds3}) becomes

\begin{equation} \label{ds5}
ds^2=\frac{1}{\left(1+Mr^2\right)^2}
\left[-d\tau^2
+ R^2 \,\left(dr^2+r^2\, d\Omega^2\right)
\right]\, ,
\end{equation}
where $M= b/a \, $
and $R=1/|a|$. This metric is conformal to FLRW, and the conformal factor
approaches unity when $M\to 0$.

\section{Constant adiabatic index}
When $\gamma$ is a constant different from $2/3$, 
the general solution of (\ref{x^2}) and (\ref{x^1}) 
becomes

\begin{equation} \label{bt}
b(t)=K\Delta t^{\frac{2}{2-3\gamma}} \, ,
\end{equation}

\begin{equation} \label{at}
a(t)=C_1 \Delta t^{-\frac{2}{3\gamma-2}}+
C_2 \Delta t^{-\frac{3\gamma}{3\gamma-2}}
-\frac{1}{3}K \Delta t^{6\frac{\gamma-1}{3\gamma-2}} \, ,
\end{equation}
thereby
\begin{equation} \label{Mt}
M(t)=K\left(C_1+C_2\Delta t^{-1}-\frac{1}{3}K\Delta t^2\right)^{-1}.
\end{equation}
Whereas for $\gamma =2/3$, we get

\be
\n{b2/3}
b(t)=K{\mbox e}^{C\Delta t} ,
\ee

\be
\n{a2/3}
a(t)=\left(C_1+C_2\Delta t-K\Delta t^2\right){\mbox e}^{C\Delta t},
\ee 
and
\be
\n{m2/3}
M(t)=K\left(C_1+C_2\Delta t-K\Delta t^2\right)^{-1},
\ee
where $K$, $C$, $C_1$ and $C_2$ are arbitrary integration constants,
and $\, \Delta t = t-t_{0}$ with $\, t_{0}$ some initial time.
Inserting (\ref{at}), (\ref{bt}), (\ref{b2/3}) and (\ref{a2/3}) in
(\ref{x^0}), (\ref{03c}) and (\ref{lambdact}) we get

\begin{equation} \label{lambdat}
\Lambda(t)=-\frac{8}{3}K^2\Delta t^{2\frac{3\gamma-4}{3\gamma-2}}+
12C_1K\Delta t^{-\frac{4}{3\gamma-2}}+16C_2K\Delta t^{-\frac{3\gamma+2}
{3\gamma-2}}+ 3C_2^2\Delta t^{-4\frac{3\gamma-1}{3\gamma-2}} \, ,
\end{equation}
and
\begin{equation} \label{qt}
q_{r}(r,t)=\frac{8Kr\Delta t^{-\frac{3\gamma}{3\gamma-2}}}
{9\left(3\gamma-2\right)}
\left[3C_2\Delta t^{-\frac{3\gamma}{3\gamma-2}}+
3\left(C_1+Kr^2\right)\Delta t^{-\frac{2}{3\gamma-2}}-
K\Delta t^{6\frac{\gamma-1}{3\gamma-2}}\right]^2 \, ,
\end{equation} 
for $\gamma\ne 2/3$, and
\be
\n{l2/3}
\Lambda(t)=\left(3C_2^2+12KC_1\right){\mbox e}^{2C\Delta t} \, ,
\ee

\be
\n{q2/3}
q_{r}(r,t)=-4Kr\left(C_1+C_2\Delta t+K(r^2-\Delta t^2)\right)^2
\,{\mbox e}^{3C\Delta t} \, ,
\ee
for $\gamma =2/3$. In the following we analyze the asymptotic
behavior of these solutions in the limits for $t$ such that $M\to 0$. 
We note by passing that in the  $M\to 0$ limit the time coordinate 
$\tau$ becomes the cosmological time, as can be seen from (\ref{ds5}). 
Two alternatives of asymptotically expanding
universes appear depending on the map between $t$ and $\tau$
-equations (\ref{taua1}), (\ref{d1/3}) and (\ref{taua2}), 
(\ref{tau2/3}) below.

\bigskip
\underline{Case $\Delta t\to 0$}

\bigskip

In this limit we obtain

\begin{equation} \label{aa1}
a\simeq C_2\Delta t^{\frac{3\gamma}{2-3\gamma}} \, ,
\end{equation}

\begin{equation} \label{taua1}
\Delta\tau\simeq\frac{2-3\gamma}{2C_2\left(1-3\gamma\right)}
\Delta t^{2\frac{1-3\gamma}{2-3\gamma}} \qquad   (\gamma\ne 1/3),
\end{equation}

\be
\n{d1/3}
\Delta\tau\simeq\frac{1}{C_2}\,{\mbox ln}\Delta t  \qquad (\gamma =1/3),
\ee

\begin{equation} \label{Ra1}
R(\tau)\simeq \frac{1}{\mid C_{2}\mid}\left[
\frac{2 C_{2} \left(1-3\gamma\right)}{2-3\gamma}
\Delta\tau\right]^{\frac{3\gamma}{2\left(3\gamma-1\right)}}
\qquad   (\gamma\ne 1/3),
\end{equation}

\be
\n{R1/3}
R(\tau)\simeq\frac{1}{\mid C_{2}\mid }\,{\mbox e}^{- C_{2} \Delta\tau}
\qquad   (\gamma = 1/3),
\ee

\begin{equation} \label{lambdaa1}
\Lambda(\tau)\simeq \frac{3\left(2-3\gamma\right)^2}
{4\left(1-3\gamma\right)^2\Delta\tau^2}      \qquad   (\gamma\ne 1/3),
\end{equation}

\be
\n{l1/3}
\Lambda\simeq 3C_2^2    \qquad   (\gamma= 1/3),
\ee

\begin{equation} \label{qa1}
q_{r}(r,\tau)\simeq \frac{8KC_2}{3\gamma-2}\,\,r
\left[\frac{2C_2\left(1-3\gamma\right)}{2-3\gamma}\Delta\tau\right]
^{\frac{9\gamma}{2\left(1-3\gamma\right)}}  \qquad   (\gamma\ne 1/3),
\end{equation}

\be
\n{q1/3}
q_{r}(r,\tau)\simeq -8KC_2^2r{\mbox e}^{3C_2\Delta\tau}  \qquad   (\gamma= 1/3).
\ee
When $1/3<\gamma<2/3$ we have, for large cosmological time $\tau$,
an accelerating universe that homogenizes with vanishing cosmological 
term and heat flow. In this stage we may define an asymptotic
adiabatic index by equating the the exponent in equation 
(\ref{Ra1}) to $2/(3\gamma_{\rm asy})$, i.e.
$\gamma=4/3\left(4-3\gamma_{\rm asy}\right)$.
In this range one has $1<\gamma_{\rm asy}<\infty$ leading to a final
power-law expansion era. For $\gamma=1/3$ the map between $t$ and $\tau$
changes and we have an asymptotically de Sitter universe with finite 
limit cosmological term. For the remaining values of $\gamma$ the
universe begins at a homogeneous singularity with a divergent cosmological
term. When $\gamma<1/3$, the heat flux asymptotically vanishes 
near the singularity, while for $\gamma>2/3$ it diverges.

\bigskip

\underline{Case $\Delta t\to\infty$}

\bigskip

In this limit we obtain
\begin{equation} \label{aa2}
a\simeq -\frac{1}{3}K \Delta t^{6\frac{\gamma-1}{3\gamma-2}},
\end{equation}

\begin{equation} \label{taua2}
\Delta\tau\simeq \frac{3}{K}\frac{2-3\gamma}{4-3\gamma}
\Delta t^{\frac{-4+3\gamma}{2-3\gamma}},
\end{equation}

\begin{equation} \label{Ra2}
R(\tau)\simeq -\frac{3}{K}\left[\frac{K\left(4-3\gamma\right)}
{3\left(2-3\gamma\right)}\Delta\tau\right]^
{\frac{6\left(1-\gamma\right)}{4-3\gamma}},
\end{equation}

\begin{equation} \label{lambdaa2}
\Lambda(\tau)\simeq -\frac{24\left(2-3\gamma\right)^2}{\left(4-3\gamma\right)^2
\Delta\tau^2},
\end{equation}

\begin{equation} \label{qa2}
q_{r}(r,\tau)\simeq -24\frac{\left(2-3\gamma\right)^2}
{\left(4-3\gamma\right)^3}\frac{r}{\Delta\tau^3},
\end{equation}
for $\gamma\ne 2/3$, and
\be
\n{a2/31}
a\simeq -K\Delta t^2\,{\mbox e}^{C\Delta t},
\ee

\be
\n{tau2/3}
\Delta\tau\simeq \,{\mbox e}^{-C\Delta t},
\ee

\be
\n{R2/3}
R(\tau)\simeq -C\Delta\tau,
\ee

\be
\n{l2/31}
\Lambda(\tau)\simeq\frac{\left(3C_2^2+12KC_1\right)}{\Delta\tau^2},
\ee

\be
\n{q2/31}
q_{r}(r,\tau)\simeq-\frac{4K^3\,r}{C^3\Delta\tau^3}\left(
{\ln}\Delta\tau\right)^4,
\ee
for $\gamma =2/3$.

In this case the asymptotic adiabatic index is given by
$\gamma=(9\gamma_{\rm asy}-4)/(9\gamma_{\rm asy}-3)$ and
we have two alternatives for asymptotically expanding universes. When
$2/3\le\gamma\le 1$ the universe  homogenizes for large cosmological
time with vanishing cosmological term and heat flow. We note that even
though an asymptotic negative cosmological term occurs, 
the universe ends
in a power-law evolution $R \propto \Delta \tau^{\alpha}$
with $0<\alpha<1$. When $\gamma=1$, the late time evolution changes to an
asymptotically Minkowski stage. For $1<\gamma<4/3$ the universe
starts homogeneously in the remote past with a vanishing scale factor,
cosmological term and heat flow. For the remaining values of $\gamma$
the universe begins at a homogeneous singularity with a divergent
cosmological term.

An exact solution with explicit dependence on the asymptotic cosmological time
$\tau$ can be found
when the integration constants $C_{1}$ and $C_{2}$ vanish. In such
a case each approximate expression (\ref{aa2})-(\ref{qa2}) 
becomes an equality and the metric reduces to

\begin{equation}
\label{metric}
ds^2=\frac{1}{\left(1+m\Delta\tau^{2\frac{2-3\gamma}{4-3\gamma}}\,r^2\right)^
2}\left[-d\tau^2+\Delta\tau^{12\frac{1-\gamma}{4-3\gamma}}\left(\ d\,r^2  +
  r^2\,d\Omega^2
\right)\right],
\end{equation}

\noindent
where $m$ is a redefinition of the old integration constant $K$, the
adiabatic index $\gamma$ and $r_0$. The last constant was introduced
by scaling the radial coordinate $r\to r_{0}\, r$.

Proposed varieties of soft matter with $\gamma<1$ include cosmic string
networks \cite{gott}, ``K--matter" \cite{kolb} and quantum zero-point field
\cite{wesson}, all with $\gamma=2/3$, as well as ``QCDM" (for ``unknown cold
dark matter") with $\simeq 0.4 $ \cite{turner1} and the quintessence (or
Q--component) with $0<\gamma<1$ \cite{caldwell}. Fluids with values of
$\gamma$ less than $2/3$ may be termed ``inflationary matter". 
Equation (\ref{Ra1}) shows an evolution that corresponds to 
this kind of matter.

These results illustrates how homogeneization of a universe
dominated by matter that has negative pressure in the present
era may have occurred. A smooth unclustered dark matter 
component with negative presure could reconcile a flat, 
or nearly flat, universe with a density in clustered 
matter well below the critical value, and moreover 
explain the recent high redshift supernovae 
data suggesting that the universe is currently under an 
accelerated expansion \cite{perlmutter1}, 
\cite{riess}.
For a perfect fluid negative presure leads to
instabilities that are most severe on the shortest 
scales. However, if instead the dark matter is a solid, 
with an elastic resistance to pure shear
deformations, an equation of state with negative 
presure can avoid these short wavelength 
instabilities. Such a solid may arise as the result of different
kinds of microphysics. Two possible candidates for a solid dark matter
component are a frustrated network of non-Abelian cosmic strings or a
frustrated network of domain walls. If these networks settle down to an
equilibrium configuration that gets carried along and stretched by the Hubble
flow, equations of state result with $\gamma=1/3$ and $\gamma=2/3$,
respectively. One expects the sound speeds for the solid dark matter component
to comprise an appreciable fraction of the light speed. Therefore, the
solid dark matter does not cluster, except on the very largest scales,
accessible only through observing the CBR anisotropy at 
large angles  \cite{bucher}.

\section{Asymptotic evolution to a quintessence--dominated era}

As a first stage of towards more general scenarios
with a slowly time--varying $\gamma$, we will explore a model 
that evolves towards an asymptotic FLRW regime
dominated by Q--matter (i.e. the scalar field). We will 
show that this system approaches to the
constant $\gamma$ solutions for large times found  above.
In this regime equations (\ref{00}) and 
(\ref{KG}) become

\begin{equation} \label{00RW}
3H^2\simeq\rho_f+\frac{1}{2}\dot\phi^2+V\left(\phi\right)+\Lambda \, ,
\end{equation}

\begin{equation} \label{KGRW}
\ddot\phi+3 H\dot\phi+ \frac{dV\left(\phi\right)}{d\phi}\simeq 0 ,
\end{equation}
where $H=\dot R/R$ and a dot means $d/d\tau$ in this section.
>From these equations
and (\ref{prhophi}) it follows
\be
\n{hp}
\dot H=-\frac{1}{2}\dot\phi^2-\frac{1}{2}\gamma_f\rho_f+
\frac{\dot\Lambda}{6H} \, ,
\ee

\noindent
and

\begin{equation} \label{gammaphi}
\gamma_{\phi}=\frac{\dot\phi^2}{\frac{1}{2}\dot\phi^2+V(\phi)}.
\end{equation}

In last section we found that the general asymptotic  solution
for the scale factor $R(\tau)\propto\Delta\tau^\alpha$ has the power--law
behaviors (\ref{Ra1}), (\ref{Ra2}) for any value of the effective adiabatic
index $\gamma$. Then, using these expressions and (\ref{lambdaa1}) and
(\ref{lambdaa2}) together with (\ref{00RW}) and (\ref{hp}) in
(\ref{gammaphi}), the latter becomes

\be
\label{gammaphi2}
\gamma_{\phi}=\frac{2}{3\alpha}
\frac{1-\frac{3\alpha\gamma_f\rho_f\Delta\tau^2}{2(3\alpha^2-\beta)}}
{1-\frac{\rho_f\Delta\tau^2}{3\alpha^2-\beta}} \, ,
\ee

\noindent
where

\begin{equation} 
\label{alpha1}
\alpha=\frac{3\gamma}{2\left(3\gamma-1\right)},\qquad
\beta=\frac{3\left(2-3\gamma\right)^2}
{4\left(1-3\gamma\right)^2},\qquad
(\frac{1}{3}<\gamma<\frac{2}{3}) \, ,
\end{equation}

\begin{equation} \label{alpha2}
\alpha=\frac{6\left(1-\gamma\right)}{4-3\gamma}, \qquad
\beta=-\frac{24\left(2-3\gamma\right)^2}{\left(4-3\gamma\right)^2},
\qquad
({\textstyle{2\over 3}}<\gamma< {\textstyle{1}}).
\end{equation}

To investigate the asymptotic limit in which the energy of the scalar field
dominates over the contribution of the perfect fluid we assume that
$3\alpha\gamma_f>2$. In this regime the two terms in (\ref{gammaphi2})
proportional to the energy density $\rho_f$ are positive and neglegible.
The adiabatic scalar field index can be approximated by
\be
\n{gammaphi3}
\gamma_ {\phi}\simeq\frac{2}{3\alpha}
\left[1+\left(1-\frac{3\gamma_f}{2}\sigma
\right)\right],
\ee
where $\sigma=\rho_f/\rho_{\phi}\ll 1$.
Inserting these equations in (\ref{gamma}) we obtain the
first correction to the effective adiabatic index for the solutions
(\ref{alpha1})

\be
\n{gamma1}
\gamma\simeq\frac{2}{3}\left[1\pm\sqrt{\left(3\gamma_f-2\right)\sigma}\right]
\ee
and $\gamma\simeq 2/3+\gamma_f\sigma/2$ for the solution (\ref{alpha2}).
The negative branch of (\ref{gamma1}) yields a consistent
asymptotic solution
for the range ${\textstyle{1\over 3}}<\gamma<{\textstyle{2\over 3}}$. 
We note that this solution
describes a deflationary stage with a limiting exponent $\alpha=1$.

Oftenly power-law evolution of the scale factor
is associated with logarithmic dependence of the scalar field on proper time
\cite{Chi98}.
Thus, assuming that $\phi(\tau)\simeq C\ln \tau$ with the constant $C$ to be
determined by the system of equations (\ref{00RW}) and (\ref{KGRW}),
and  using these expressions together with (\ref{lambdaa1}) and 
(\ref{lambdaa2}) in (\ref{00RW}) and (\ref{KGRW}) it follows

\begin{equation} \label{00a}
\frac{3\alpha^2}{\tau^2}\simeq
\frac{C^2}{2\tau^2}+V+\frac{\beta}{\tau^2} \, ,
\end{equation}

\begin{equation} \label{KGa}
\frac{\left(3\alpha-1\right)C}{\tau^2}+\frac{dV}{d\phi}\simeq 0 \, .
\end{equation}

{}From (\ref{00a}), (\ref{KGa}) we obtain the leading term of 
$V(\phi)$ for large $\phi$

\begin{equation} \label{Va}
V\left(\phi\right)\simeq V_0 e^{-A\phi}
\end{equation}

\noindent
and

\begin{equation} \label{A1}
A^2=3\gamma, \qquad
V_0={\frac {3\,\gamma+2}{3\gamma\,\left (3\,\gamma-1\right )}},
\end{equation}
for ${\textstyle{1\over 3}}<\gamma<{\textstyle{2\over 3}}$, while

\begin{equation} \label{A2}
A^2=3\,{\frac {\left (-4+3\,\gamma\right )\left (-1+\gamma\right )}{17-42
\,\gamma+27\,{\gamma}^{2}}},
\qquad
V_0={\frac {2\left (-14+15\,\gamma\right )\left (17-42\,\gamma+27\,{
\gamma}^{2}\right )}{3\left (-1+\gamma\right )\left (-4+3\,\gamma
\right )^{2}}} \, ,
\end{equation}
for ${\textstyle{2\over 3}}<\gamma<1$.
Inserting the dominant value of the effective adiabatic index
in (\ref{A1}), (\ref{A2}) 
we find $A^2=2$, $V_0=2$ and $C=1/\sqrt{2}$.

The models considered in this section are
based on the notion of ``late time dominating field" (LTDF),
a form of quintessence in which the field $\phi$ rolls down a
potential $V(\phi)$ according to an attractor solution to the
equations of motion.
This solution is an attractor since for a very wide range of
initial values for $M$, $\phi$ and $\dot{\phi}$ it rapidly approaches a
common evolutionary path, i.e. the late behavior is
insensitive to the initial conditions.   
This model has an advantage similar to
inflation in that for a wide range of initial conditions
the universe is driven to the same final evolution.
The ratio $\sigma$ of the background fluid to the field energy
changes steadily as $\phi$ proceeds down its path.
This is desirable because in that way the Q-matter 
ultimately dominates the energy density and drives the universe
toward an accelerated expansion. 

Recently Ferreira and Joyce \cite{Ferreira} proposed a model 
based on an exponential potential. Their self-adjusting 
solutions are attractors and
$\Omega_\phi$ remains constant for a constant background 
equation of state as $\gamma_{\phi} = \gamma_{f}$.
($\Omega_\phi$ changes slightly when the universe shifts 
from radiation--dominated  to matter--dominated expansion).
This means, for example, 
that $\Omega_{\phi}$ is constant throughout the matter--dominated
epoch.  For a constant $\Omega_{\phi}$
to satisfy the structure formation constraints requires 
$\Omega_{\phi} < 0.2$  and $\Omega_{f} > 0.8$. This however
runs into conflict with the current best estimates of $\Omega_{f}$ and
produces a decelerating universe at variance with recent supernovae 
observations \cite{perlmutter1}, \cite{riess}.

By contrast, our LTDF model only requires that the potential has an
asymptotic exponential shape for large $\phi$.
So, the interesting and significant features of our
model are: (a) like the self-adjusting case, a wide range
of initial conditions are drawn towards a common evolution; 
however, (b) the LTDF solutions do not ``self-adjust" to the
background equation of state but rather,
maintain some finite difference in the equation of state
such that the  field energy  eventually
dominates and the universe enters a period of acceleration. 
Compared to the self-adjusting model, ours does not require any
additional parameter and allows a much wider range of potentials, 
provided they have an exponential tail. LTDF solutions exist 
for a very wide class of potentials. The energy density
of the field decreases as $R^{-2}$, and this power--law 
remains unchanged in every epoch of the universe when the 
background expansion turns from radiation- to matter- to 
quintessence-dominated.
The value of $\gamma_\phi$ differs from the background equation 
of state such that the value of $\Omega_\phi$ increases as the 
universe ages. Hence, $\Omega_\phi$ grows to order unity late 
with time.

\section{General solution for a time--dependent adiabatic index}

A time-varying $\gamma$ is very natural because
different matter components redshift at various 
rates. Then the question arises whether the conditions leading to
homogeneization we have found with constant $\gamma$ also hold
when $\gamma$ varies with time.
The rationale behind this approach is the following. 
When  different components enter the stress-energy
tensor of the cosmic medium, it is natural
to expect that each epoch is dominated by
the energy density of just a 
single component with a constant, or nearly constant, 
adiabatic index, say $\gamma_{01}$. The others components 
can be regarded as small perturbations. As the universe 
expands, sooner or later, the component 
with the adiabatic index inmediately lower than 
$\gamma_{01}$ (say $\gamma_{02})$ takes over,
and a new epoch begins. 

We start by giving the general solution to equations
(\ref{x^2}) and (\ref{x^1}) when $\gamma$ is a 
function of time

\begin{equation} \label{bw}
b=\exp{\left(\int dt \,w\right)}\, ,
\end{equation}

\begin{equation} \label{aw}
a=-2\exp{\left(\int dt \,w\right)}\int dt \,w^2
\int \frac{dt}{w^2} \, ,
\end{equation}
where

\begin{equation} \label{wt}
w=\frac{2}{\int dt \left(2-3\gamma\right)} \, ,
\end{equation}
provided $\gamma\neq 2/3$.
Inserting (\ref{aw}) and (\ref{bw}) in (\ref{x^0}), (\ref{03c}) and
(\ref{lambdact}) it follows

$$
\Lambda=4K^2\exp{\left(2\int dt \,w\right)}\left[3w^4\left(
\int \frac{dt}{w^2}\right)^2+
\frac{4w^3}{\gamma}\left(\int dt \,w^2\int \frac{dt}{w^2}
\right)\int \frac{dt}{w^2}\right.
$$
\begin{equation} \label{lambdaw}
\left. + \frac{2w^2}{\gamma}
\left(\int dt \,w^2\int \frac{dt}{w^2}\right)^2-
6\int dt \,w^2\int \frac{dt}{w^2}\right] \, ,
\end{equation}

\begin{equation} \label{qw}
q_{r}(r,t)=-4K^3rw\exp{\left(3\int dt \,w\right)}
\left(-2\int dt \,w^2\int \frac{dt}{w^2}+r^2\right)^{2} .
\end{equation}

Though we have reduced the general coupled system of equations to quadratures,
it is very involved to obtain the solution in closed form except for 
constant $\gamma$. This is why next subsection focus on
approximated solutions assuming that $\gamma(t)$ admits an 
analytic expansion.

\subsection{Homogeneization with a varying adiabatic index}

The  transition previously described  from one epoch to the next, assumed to
be gentle, may be modeled by an homographic function $\gamma(t)
=(At+B)/(Ct+D)$ where $A, B, C, \mbox{and} \, D$ are constants \cite{mendez}.
With this in mind, we next investigate two kind of behaviors for the fluid
components that lead to a final homogeneous stage characterized by $M(t)\to
0$. These can be associated with two asymptotic evolutions of $\gamma(t)$.

\bigskip
\begin{enumerate}
\item
In the limit $t\to 0$ we assume that $\gamma(t)$ has a 
Taylor expansion

\begin{equation} \label{gammaT0}
\gamma(t) = \gamma_{01}+\gamma_{11} t + {\cal O}(t^2) ,
\qquad \gamma_{01} = \frac{B}{D} \, , 
\quad \gamma_{11} = \frac{A}{D} - \frac{B}{D^{2}} \, ,
\end{equation}
with $\gamma_{01}= B/D$ and $\gamma_{11} = (A D^{-1} - B D^{-2})$. 
Then, for $\gamma_{01}\neq 2/3$ we find that

\begin{equation} \label{MT0}
M(t)=\frac{K\,t}{C_2}\left[1-\frac{3\gamma_{11}t}{2-3\gamma_{01}}\ln t+
{\cal O}\left(t^2\right)\right] \, ,
\end{equation}
and that the first corrections to equations
(\ref{aa1}) and (\ref{qa1}) vanish as $t\ln t$, while the first correction 
to (\ref{lambdaa1}) is of higher order.

\item
In the limit $t\to \infty$ we assume for  $\gamma(t)$ the
expansion

\begin{equation} \label{gammaTinf}
\gamma(t)=\gamma_{02}+\frac{\gamma_{12}}{t} + {\cal O}(t^{-2}),
\qquad \gamma_{02} = A/C \, , \quad 
\gamma_{12}= \frac{B}{C} - \frac{A}{C^{2}} \, ,
\end{equation}
obtaining for $\gamma_{02}\neq 2/3$

\begin{equation} \label{MLinf}
M(t)=-\frac{3}{t^2}\left[1+\frac{6\gamma_{12}}{2-3\gamma_{02}}
\frac{\ln t}{t}+{\cal O}(t^{-1})\right].
\end{equation}
In this case the first corrections to equations
(\ref{aa2}), (\ref{lambdaa2}) and (\ref{qa2}) vanish as $\ln t/t$.
\end{enumerate}

These results show that homogeneization occurs
under the same conditions for both 
$\gamma_{0}$ as it does for constant $\gamma$ 
for a wide range of evolutions 
of the adiabatic index, 
provided it has a constant limit, $\gamma_{01}$ and $\gamma_{02}$
for $t\to 0$ and  $t\to\infty$, repectively, 
and is analytic about these points.

\section{Concluding remarks}

We have investigated a class of solutions of the Einstein field equations with
a variable cosmological term, heat flow and a fluid with variable adiabatic
index that includes those of Modak, Bergmann and Maiti and contains a new
exact conformally flat solution. We have also considered the contribution of a
homogeneous minimally coupled scalar field to the stress--energy tensor. The
solutions of the EKG system that lead to an asymptotic FLRW stage were
analyzed when the adiabatic index remains constant. We have found that
asymptotically expanding universes occur when $1/3<\gamma<1$ that homogenizes
for large cosmological time  with vanishing cosmological term and heat flow.
For $1/3<\gamma<2/3$ the evolution is given by (\ref{Ra1}) and corresponds to
a power-law accelerated expansion for large cosmological time $\tau$,
as follows from (\ref{taua1}). On the other hand, when
$2/3\le\gamma<1$ even though an asymptotic negative cosmological term occurs,
the universe evolves toward a decelerated expansion. The particular case
$\gamma=1/3$ leads asymptotically to a de Sitter universe with a finite limit
for $\Lambda$. We have shown that these results also apply to a time dependent
adiabatic index.

We have carried out a detailed analysis of a model in which Q--matter dominates
over cold dark matter. This LTDF solution is an attractor because, even for
large initial inhomogeneities and a wide range of initial values for $\phi$
and $\dot{\phi}$, the evolution approaches a common path. It was shown that
this model can be realized for a wide range of potentials provided they have
an exponential tail. This is quite interesting because recently, there has
been increasing activity related to scalar fields with the Liouville form
(exponential) potential. It arises as an effective potential in many theories
such as Jordan-Brans-Dicke theory, Salam-Sezgin theories and superstring
theories. Indeed, most theories undergoing dimensional reduction to an
effective four-dimensional theory, result in a linear combination of
exponential potentials, and one of these will eventually dominate. 

This work can be generalized in different directions. The more obvious
one is to allow $\Lambda$ and $\gamma$ to depend on position too, but
this seems analytically impracticable.
A less hard  possibility is
to consider non-flat spatial sections. This is interesting since
although the location of the first acoustic peak in the angular power
spectrum of the CBR suggests a flat universe,
its exact position is still uncertain \cite{silk}. Therefore 
it may well happen that the universe is open or even closed. 
Again the difficulty with the corresponding analysis is basically 
mathematic since it is doubtful that in this more general case
the EKG field equations admit analytical solutions. A seemingly 
less involved generalization is to include a bulk dissipative 
pressure in the  stress--energy tensor (\ref{2}), something rather 
natural \cite{dark}. Its effect should tend to further accelerate the 
expansion, as it lowers the total pressure of the cosmic fluid.  
The problem with this is the lack of fully realiable expressions for the 
coefficient of bulk viscosity of the fluids involved in the hydrodynamic
description. However, in this regard the situation is  no  worse
than that encountered in inflationary scenarios in which potentials
for the scalar field are frequently proposed with no obvious
physical ground.

\section*{Acknowledgments}

This work was partially supported by the Spanish Ministry 
of Education under Grant PB94-0718, and the University of 
Buenos Aires under Grant TX-93.

\end{document}